\documentclass[12pt]{article}
\usepackage{amssymb}

\def\tfrac#1#2{{\textstyle {#1 \over #2}}}%

\font\extra=msbm10 scaled \magstep1
\def\bbb #1{\hbox{{\extra #1}}}

\begin{document}

\begin{center}
{\LARGE{\bf Discrete supersymmetries of the Schr\"odinger equation
and non-local exactly solvable potentials
}}\footnote{Published in Phys. Lett. A 302 (2002) 234-241}

\vskip1cm

Boris F. Samsonov${}^1$ and
A. A. Suzko${}^2$
\vskip0.5cm
{\it
${}^1$Department of Quantum Field Theory, Tomsk State
University \\ 36 Lenin Ave., 634050 Tomsk, Russia,
e-mail: samsonov@phys.tsu.ru
\\  [1ex]
${}^2$Joint Institute for Nuclear Research, 141980 Dubna, Russia, e-mail:
suzko@cv.jinr.ru}
\end{center}
\vskip0.5cm

\begin{abstract}
\noindent
Using an isomorphism between Hilbert spaces $L^2$ and
$\ell^{\,2}$ we consider Hamiltonians which have tridiagonal
matrix representations (Jacobi matrices)
in a discrete basis and an eigenvalue
problem is reduced to solving a three term difference equation.
Technique of intertwining operators is applied
to creating new families of exactly solvable Jacobi matrices.
It is shown that
any thus obtained Jacobi matrix gives rise to a
new exactly solvable non-local potential of the
Schr\"odinger equation.
We also show that the algebraic structure underlying our approach
corresponds to supersymmetry.
Supercharge operators acting in the space $\ell^{\,2}\times \ell^{\,2} $ are
introduced which together with a matrix form of the
superhamiltonian close the simplest superalgebra.

\end{abstract}

\section{Introduction}

Supersymmetric quantum mechanics (SUSY QM) first introduced by Witten
\cite{Wit} as a simplest supersymmetric model of the quantum field theory
represents now a fast progressing field of modern theoretical
physics (see recent books \cite{books}).
As it has been understood after the work by Andrianov et al
\cite{Andr}, the SUSY QM may be considered as another formulation of
Darboux transformation method well-known in mathematics from the
original paper by Darboux \cite{Darb}, book by Ince \cite{Ince}, and book
by Matveev and Salle \cite{MatvSalle} where this method is widely used in
the context of the soliton theory.
An essential ingredient of
the method is a particular choice of
a transformation operator
(or intertwining operator \cite{Levitan})
in the form of a differential operator
(see e.g. \cite{BSTMF}) which intertwines
two Hamiltonians and relates their eigenfunctions.
This approach being applied in quantum
theory allows one to generate
a huge family of  exactly solvable local potentials starting
with a given exactly solvable local potential.

The technique of transformation operators,
first proposed by Delsart \cite{Delsart}
(he calls them transmutation operators, see also e.g.
\cite{Levitan})
in connection with generalized
translation operators,
is essentially based on intertwining relations,
which have an universal character \cite{shabat}.
It is applicable
not only to differential equations (like the Schr\"odinger
equation) but also to difference equations
\cite{Crist,geron,MatvSalle}
which, in the simplest case,
represent eigenvalue problems for Jacobi matrices.
Intertwining operators
are widely used in the context of soliton solutions of nonlinear lattice
equations  (see \cite{MatvSalle} and references therein).

It is known that spectral problems for finite-difference
(or discrete) Schr\"odinger equations
have much in common with the theory of orthogonal polynomials
\cite{berez}.
The Darboux trans\-for\-mation
method gives new insight to this theory and is very fruitful in
establishing new properties and deeper understanding some known
relations \cite{SpiridZhed}.
As it was shown in \cite{shabat},
the continuous Darboux transformation is
intimately related with quantum inverse scattering problem.
A discrete version of this problem
has been considered in ~\cite{case}
(for a review see e.g, \cite{book})
but its relationship with discrete
Darboux transformations
is unknown to the authors.
We note that considerable attention has been drown to the
non-Hermitian form of the problem
\cite{MatvSalle,Crist,geron,SpiridZhed},
while in quantum mechanics one usually uses
self-adjoint Hamiltonians.
The study of the Hermitian problem has been just started
\cite{Asuz}.
We thus notice that the discrete version of the
Darboux transform though ascending to
Christoffel \cite{Crist}
 and Geronimus ~\cite{geron} works
is developed much less than the continuous one.

This paper is aimed to partially fill in this gap.
We apply the intertwining relation to a discrete eigenvalue
problem for a Jacobi matrix.
This allows us to generate a family of new exactly solvable Jacobi
matrices.
Eigenvalue problems for these matrices
naturally appear in quantum mechanics
when a discrete basis is used for solving the Schr\"odinger
equation.
In particular, there exist
exactly solvable Hamiltonians represented by Jacobi matrices
\cite{YamFish}.
We show that new Jacobi matrices define new exactly
solvable potentials of a non-local nature.
Moreover, when two lattice eigenvalue problems
are written as a single matrix problem, nilpotent difference supercharges
may be introduced to map the solutions of these problems to each
other. We also show that these supercharges
together with a difference superhamiltonian close the
simplest superalgebra. We associate these supercharges with
discrete supersymmetries of the Schr\"odinger equation.

\section{Intertwining in $\ell ^{\,2}$-space}

Consider self-adjoint Hamiltonians $h_0$ and $h_1$ defined in a
Hilbert space $H=L^2({\Bbb R})$. We do not exclude the presence of
a continuum spectrum and if necessary consider them as defined in
a wider space $H_-$ of  linear functionals over $H_+$, $H_+\subset
H\subset H_-$\,, where $H_-$ and $H_+$ are
the Hilbert-Schmidt equipment of the space $H$
(so called Gel'fand triplet, see e.g. \cite{GSh})
without especially mentioning it.
Let $|\psi_E\rangle $ and  $|\widetilde{\psi}_E\rangle $ be eigenkets of
$h_0$ and $h_1$ respectively with an eigenvalue $E$,
$h_0|\psi_E\rangle =E|\psi_E\rangle $,
$h_1|\widetilde\psi_E\rangle =E|\widetilde\psi_E\rangle $,
and $|n\rangle $ be a basis (orthogonal or not) in $H$.
The basis $|n\rangle $ is supposed to be such that
the action of both $h_0$ and $h_1$ takes the form of three term
relations:
\begin{equation}\label{h00n}
h_{\,0}\,|n\rangle =a_n|n-1\rangle +a_{n+1}|n+1\rangle
+{q}_n|n\rangle ~\,
\end{equation}
\begin{equation}\label{h11n}
h_1\,|n\rangle =\widetilde{a}_n|n-1\rangle +
\widetilde{a}_{n+1}|n+1\rangle +\widetilde{q}_n|n\rangle \,.
\end{equation}
We assume that $n$ takes positive integers including zero,
$n=0,1,2,\ldots $ and $a_0=\widetilde{a}_0=0$.

Denote by $\psi_n=\psi_n(E)$ and  $\widetilde{\psi}_n=\widetilde{\psi}_n(E)$
the Fourier coefficients
of $|\psi_E\rangle $ and  $|\widetilde{\psi}_E\rangle $
in terms of the basis $|n\rangle $ respectively:
\begin{equation}\label{psiE}
|\psi_E\rangle =\sum_n \psi_n|n\rangle \,,\quad
|\widetilde{\psi}_E\rangle =\sum_n \widetilde{\psi}_n|n\rangle \,.
\end{equation}
Equations (\ref{h00n}) and (\ref{h11n}) define the action of
$h_0$ and $h_1$ on any
$|\psi\rangle =\sum c_n|n\rangle \in H$
where the sum contains a finite number of items.
(They form a finite set in $H$.)
An eigenket of a Hamiltonian belongs usually to
a wider domain which can be specified in terms of
its Fourier coefficients over a basis.
We will not discuss these mathematical subtleties here
and shall simply suppose that all Fourier series converge
and when an operator acts on a series
it may be moved through the sign of sum.
Thus, having (\ref{h00n}) and (\ref{h11n}) in mind we can act
by $h_0$ and $h_1$ on $|\psi_E\rangle $ and
$|\widetilde\psi_E\rangle $ given by (\ref{psiE})  respectively.
As a result we obtain the Fourier series
\begin{equation}
h_0|\psi_E\rangle =\sum_n (h_0\psi)_n|n\rangle \,,
\quad \label{h0m}
h_1|\widetilde{\psi}_E\rangle =\sum_n (h_1\widetilde{\psi})_n|n\rangle
\end{equation}
with the Fourier coefficients
\begin{eqnarray}
(h_0\psi)_n =a_n\psi _{n-1} +a_{n+1}\psi_{n+1} +q_n\psi_n ~,
\label{h0n}\\
(h_1\widetilde{\psi})_n =
\widetilde{a}_n\widetilde{\psi}_{n-1} +\widetilde{a}_{n+1}\widetilde{\psi}_{n+1}
 +\widetilde{q}_n\widetilde{\psi}_n\,\,.
\label{h1n}
\end{eqnarray}
Note that these quantities depend on the energy $E$ but
we do not indicate this to avoid cumbersome notations.
The eigenvalue problems for $h_0$ and $h_1$ are reduced
now to
the second order three-term finite difference equations
\begin{eqnarray}
a_n\psi _{n-1} +a_{n+1}\psi_{n+1}+q_n\psi_n=E\psi_n\,,
\label{init}
\\
\widetilde{a}_n\widetilde{\psi}_{n-1} +\widetilde{a}_{n+1}\widetilde{\psi}_{n+1} +\widetilde{q}_n\widetilde{\psi}_n=
E\widetilde{\psi}_n\label{fin}\,.
\end{eqnarray}

Note that the space of sequences $\left\{ \psi_n \right\}$ with
an appropriately defined inner product is known as the  space
$\ell ^{\,2}$ which also is a Hilbert space (if the space is not
complete, it always may be enlarged to become a Hilbert space).
If necessary this space may also be equipped with $\ell^{\,2}_\pm$,
$\ell^{\,2}_+\subset \ell^{\,2}\subset \ell^{\,2}_-$.
The meaning of $\ell^{\,2}_\pm$ is the same as $H_\pm $.
Eqs. (\ref{h0n}) and (\ref{h1n}) define matrix representations
of the Hamiltonians $h_0$ and $h_1$.

A second order finite difference equation like a second order
differential equation has two linearly independent solutions.
This means that for a fixed $E$ one has a two-dimensional
linear space of solutions to Eq.(\ref{init}) (and respectively
(\ref{fin})).
When the spectrum of a Hamiltonian is non-degenerate
and the energy is fixed, in this space
may exist a unique (up to normalization) element
that can be associated with an eigenvector of $h_0$.
For a bound state eigenvalue $E$ it may be selected from the
condition
$$\langle h_0\psi_E|h_0\psi_E\rangle =
\sum_{n,\,k} (h_0\psi )_{k}^*\,(h_0\psi )_n \langle k|n\rangle <\infty
\,.$$
We use $"{}^*"$ to distinguish complex conjugate quantities.
For a continuous spectrum this value may diverge like the
Dirac $\delta $ function,
$$\langle h_0\psi_E|h_0\psi_{E'}\rangle =
\sum_{n,k} (h_0\psi )^{*}_k(E)\cdot (h_0\psi )_n (E{\,'}\,)
\langle k|n\rangle \sim \delta (E-E{\,'}\,) \,.$$
In these equations $(h_0\psi )_n\equiv (h_0\psi )_n (E)$ are defined in
(\ref{h0n}).
Note also that the continuous spectrum may be two-fold
degenerate. In this case any solution of Eq.
(\ref{init}) has a physical meaning.
The right-hand sides of Eqs. (\ref{h0n}) and
(\ref{h1n}) define the action of the operators $h_0$ and $h_1$
in the space $\ell^{\,2}$, e.g.
$h_0\left\{\psi_n\right\}\equiv \left\{(h_0\psi )_n\right\}$.

Denote
 $\widehat{\psi}_n$
another solution of Eq. (\ref{init})
corresponding to the same energy $E$. Then excluding $q_n$ from
eigenvalue equations for $\psi_n$ and  $\widehat{\psi}_n$,
one arrives at the equation
\begin{equation}
a_n(\widehat \psi_n{\psi}_{n-1}-\widehat \psi_{n-1}{\psi}_{n})=
a_{n+1}(\widehat \psi_{n+1}{\psi}_{n}-\widehat \psi_{n}{\psi}_{n+1})
\end{equation}
which gives us a discrete analogue of the
Wronskian as a combination of $\psi_n$ and $\widehat{\psi}_n$
independent of the discrete variable $n$
\begin{equation}
W(\widehat \psi_n,{\psi}_n)=
a_n(\widehat{\psi}_n\psi_{n-1}-\widehat{\psi}_{n-1}\psi_n)=w_0=
{\rm const}(n)\,.
\end{equation}
When $w_0$ and $\psi_n$ are fixed, this equation
is a recursion relation for
$\widehat\psi_n$ which has the following solution
\begin{equation}\label{tpsin}
\widehat{\psi}_{n}=\frac{\widehat{\psi}_0}{\psi_0}\,\psi_n+
\sum_{k=1}^n\frac{w_0\psi_n}{a_k\psi_k\psi_{k-1}}~.
\end{equation}
For solutions $\psi_{n}$ and $\xi_n$ of (\ref{init})
with different eigenvalues
$E$ and $\lambda$
respectively
similar calculation gives a recursion relation
for $W(\xi_n,\psi_n)\equiv W_n$,
\begin{equation}
W_{n+1}=W_n+(\lambda -E)\xi_n\psi_n\,,
\end{equation}
which  may be easily iterated to become
\begin{equation}
W_{n+1}=W_1+(\lambda-E)\sum_{k=1}^n\xi_k\psi_k\,.
\end{equation}

Let us suppose that the eigenkets of $h_0$, $|\psi_E\rangle $, are
known, i.e. one knows solutions of Eq. (\ref{init}).
Then the eigenvalue problem for $h_1$,
i.e. the search for solutions of (\ref{fin}),
may be replaced by
the search for an intertwiner $L$ defined as
\begin{equation}\label{intertw}
Lh_0=h_1L\,.
\end{equation}
Once $L$ is found, the eigenkets $|\widetilde\psi_E\rangle $ are obtained
by applying $L$ to $|\psi _E\rangle $,
$|\widetilde\psi_E\rangle =L|\psi _E\rangle $.

Let us consider a
particular anzats for $L$ assuming that it is defined in terms
of the basis $|n\rangle $ as follows:
\begin{equation}\label{Lonn}
L|n\rangle =A_n|n-1\rangle +B_n|n\rangle \,.
\end{equation}
Then the Fourier coefficients
$\widetilde\psi_n=L\psi_n\equiv (L\psi )_n$ of
$|\widetilde\psi\rangle =L|\psi\rangle=\sum _n\widetilde\psi_n|n\rangle $
are expressed in terms of $\psi_n$:
\begin{equation}\label{Lpsin}
\widetilde\psi_n=(L\psi )_n=A_{n+1}\psi_{n+1}+B_n\psi_n\,.
\end{equation}

To determine $A_n$, $B_n$ defining $L$ and $\widetilde{a}_n$, $\widetilde{q}_n$ defining $h_1$,
we shall use the intertwining relation (\ref{intertw}).
The successive application of (\ref{h0n}) and (\ref{Lpsin}) gives
 $$L(h_0 \psi )_n= A_{n+1}(a_{n+2}\psi_{n+2}+q_{n+1}\psi_{n+1}+a_{n+1}\psi_{n})
 +B_n(a_{n+1}\psi_{n+1}+q_{n}\psi_{n}+a_{n}\psi_{n-1})$$
whereas the formulae (\ref{Lpsin}) and (\ref{h1n}) yields
$$h_1(L \psi )_n = \widetilde a_{n+1}(A_{n+2}\psi_{n+2}+B_{n+1}\psi_{n+1})
+\widetilde q_n(A_{n+1}\psi_{n+1}+B_{n}\psi_{n})+
\widetilde a_{n}(A_{n}\psi_{n}+B_{n-1}\psi_{n-1})$$
from which one finds the following system of equations:
\begin{eqnarray}
& &A_{n}a_{n+1}=A_{n+1}\widetilde{a}_{n}\,,\label{sys1}\\
& &B_na_n=B_{n-1}\widetilde{a}_n\label{sys2}\,,\\
& &A_{n+1}q_{n+1}+B_n a_{n+1}=
B_{n+1}\widetilde{a}_{n+1}+A_{n+1}\widetilde{q}_n\,,\label{sys3}\\
& &A_{n+1}a_{n+1}+B_nq_n=\widetilde{a}_nA_n+B_n\widetilde{q}_n\,.
\label{sys4}
\end{eqnarray}
Having found $\widetilde{a}_n$ from (\ref{sys2}) and replaced it in
(\ref{sys1}), one can "integrate" the latter equation to get
\begin{equation}
B_n=A\frac{a_{n+1}}{A_{n+1}}
\end{equation}
where $A$ denotes an "integration constant".
After excluding $\widetilde{q}_n$, $\widetilde{a}_n$ and $B_n$ from (\ref{sys3}) and
(\ref{sys4})  one obtains the equation for $A_n$
\begin{equation}
q_n-\frac{Aa_{n+1}}{A^2_{n+1}}a_{n+1}-\frac{A^2_n}{Aa_n}a_n=
q_{n+1}-\frac{Aa_{n+2}}{A^2_{n+2}}a_{n+2}-
\frac{A^2_{n+1}}{Aa_{n+1}}a_{n+1}\,.
\end{equation}
Clearly, both left- and right-hand sides of this
equation are independent on $n$. Then denoting this quantity by
$\lambda $ and introducing a new variable $\xi_n$
\begin{equation}
\frac{\xi_{n-1}}{\xi_n}=-\frac{A_n^2}{Aa_n}
\end{equation}
one arrives at the equation for $\xi_n$:
\begin{equation}\label{u}
\xi_{n-1}a_n+\xi_{n+1}a_{n+1}+(q_n-\lambda)\xi_n=0~.
\end{equation}
The latter equation is nothing but the initial eigenvalue
problem (\ref{init}) which is supposed to be solved. We thus
shall suppose that the sequences $\left\{\xi_n(\lambda )\right\}$
and
$\left\{\right.$$\widehat\xi_n(\lambda )$$\left.\right\}$
being linearly independent solutions of (\ref{u})
 for a fixed value of $\lambda $ are known.
 When $\lambda =E$ is a point of the
spectrum of $h_0$, there exists their linear combination which
gives the ket $|\psi_E\rangle $ in $n$-representation
for non-degenerate $E$.
For two-fold degenerate $E$ both
$\left\{\xi_n(E )\right\}$
and
$\left\{\right.$$\widehat\xi_n(E)$$\left.\right\}$
define two linearly independent kets.

Having found $A_n$ and $B_n$:
\begin{equation}\label{AnBn}
A_n=[-Aa_n\xi_{n-1}/\xi_n]^{1/2}\,,\quad
B_n=-[-Aa_{n+1}\xi_{n+1}/\xi_n]^{1/2}~,
\end{equation}
one gets solutions $\widetilde \psi_n(E)$ of the transformed equation (\ref{fin})
\begin{equation}\label{vfin}
\widetilde \psi_n(E) =(L\psi )_n(E)=
\left[ \frac{-Aa_{n+1}}{\xi_n\xi_{n+1}} \right]^{1/2}
\left( \xi_n\psi_{n+1}(E)-\xi_{n+1}\psi_n(E) \right)\,,
\end{equation}
\begin{equation}
(h_1\widetilde \psi \,)_n(E)=E\widetilde\psi_n(E)~.
\end{equation}
From (\ref{sys2}) one obtains the elements $\widetilde{a}_n$
\begin{equation}\label{bn}
\widetilde{a}_n=\frac{1}{\xi_n}\left[ a_na_{n+1}\xi_{n+1}\xi_{n-1} \right]^{1/2}
\end{equation}
and from (\ref{sys3}) and (\ref{sys4}) one has two equivalent
expressions for $\widetilde{q}_n$:
\begin{equation}\label{qn}
\widetilde{q}_n=q_n+a_n\frac{\xi_{n-1}}{\xi_n}-a_{n+1}\frac{\xi_n}{\xi_{n+1}}=
q_{n+1}-a_{n+1}\frac{\xi_{n+1}}{\xi_n}+a_{n+2}\frac{\xi_{n+2}}{\xi_{n+1}}~.
\end{equation}

The formula (\ref{vfin}) defines the sequence
$\left\{\right.$$\widetilde\psi_n(E)$$\left.\right\}$
for any $E\ne \lambda $. When $E=\lambda $, it gives zero, $(L\xi)_n=0$,
but we can find a solution of Eq. (\ref{fin}) at $E=\lambda $ by
acting with $L$ to another solution of Eq. (\ref{u}),
$\widehat\xi_n$,
which may be found using (\ref{tpsin}) with the replacement
$\psi\to \xi$. Thus we have the sequence $\left\{\eta_n\right\}$ given by
\begin{equation}\label{vn}
\eta_n=(L\widehat\xi \,)_n=[a_{n+1}\,\xi_n\,\xi_{n+1}]^{-1/2}\,,\quad
(h_1\eta )_n=\lambda \eta_n
\end{equation}
where inessential constant $w_0$ is omitted.
Another solution $\widetilde \eta_n$ of Eq. (\ref{fin})
at $E=\lambda $ may be found by appropriately changed
formula (\ref{tpsin})
\begin{equation}\label{tvn}
\widehat \eta_n=\eta_n\,[\,\widehat \eta_0/\eta_0+w_0\sum_{k=1}^n\xi_{\,k}^2\,]\,,\quad
(h_1\widehat \eta \,)_n=\lambda \widehat \eta_n~.
\end{equation}
So, we know all solutions of the transformed discrete equation
(\ref{fin}) with $\widetilde{a}_n$ and $\widetilde{q}_n$ given by (\ref{bn}) and
(\ref{qn}) provided all solutions of (\ref{init}) are known.
When $E\ne \lambda$, they are given by (\ref{vfin}) and
for $E=\lambda $ the formulae (\ref{vn}) and (\ref{tvn}) should be
used.
Note that the fact that the sequence
$\left\{\right.$$\widetilde\psi_n$$\left.\right\}
=\left\{(L\psi)_n\right\}$
satisfies the difference equation (\ref{fin})
means that in the space $\ell^{\,2}$ the intertwining relation
$Lh_0=h_1L$ holds where the action of $h_0$ on a sequence
is defined by (\ref{h0n}) and the action of $L$ is defined by
(\ref{Lpsin}) with $A_n$ and $B_n$ from (\ref{AnBn}).

The most popular bases in quantum mechanics are orthonormal
bases, $\langle n|n'\rangle =\delta_{n,n'}$. For such a basis
the action of the operator $L^+$, Hermitian conjugate to $L$ with
respect to the inner product, on any finite vector
$|\widetilde\psi\,\rangle =\sum _n\widetilde\psi_n|n\rangle$ where the sum contains
only a finite number of items can be easily found. For this
purpose one first notices that for an $L$ given by (\ref{Lonn})
one has
$\langle n'|Ln\rangle =A_n\delta _{n',n-1}+B_n\delta_{n',n}=
A_{n'+1}\delta_{n'+1,n}+B_n\delta _{n',n}$. Now from the
condition $\langle n'|Ln\rangle =\langle L^+n'|n\rangle $
one deduces that
$L^+|n\rangle =A_{n+1}^{*}|n+1\rangle +B_{n}^*|n\rangle$
from which it follows that
\begin{equation}\label{Ldag}
L^+|\widetilde\psi\,\rangle =\sum_n(L^+\widetilde\psi \,)_n|n\rangle \,,\quad
(L^+\widetilde\psi \,)_n= A_n^*\widetilde\psi_{n-1}+B_n^*\widetilde\psi_n
\end{equation}
with $A_n$ and $B_n$ given in (\ref{AnBn}).
Since the Hamiltonians $h_0$ and $h_1$ are self-adjoint, the
conjugate operator $L^+$ satisfies the conjugate intertwining
relation $h_0L^+=L^+h_1$ which means that $L^+$ realizes the
transformation in the opposite direction, from solutions of
(\ref{fin}) to solutions of (\ref{init}) but it is not inverse
to $L$. We shall find
the superpositions $L^+L$ and $LL^+$ in the next section.

For a self-adjoint Hamiltonian it is natural to suppose that
$a_n$ and ${q}_n$ are real for all $n=0,1,\ldots $. This implies
that for a real $\lambda $ all $\xi _n$ and $\widehat \xi_n$ may be chosen
real which from now on will be assumed. This means that the
variables $A_n$ and $B_n$ defined by (\ref{AnBn}) may become
complex only owing to the constant $-A$.
Since the operator $L^+$ is well-defined for any finite
element from $\ell^{\,2}$, it is not difficult to convince ourselves
that $(L^+\eta )_n=0$ and
$(L^+\widehat \eta \,)_n=-w_0\sqrt{{-A}^{*}\vphantom{A^{A^a}}\,}\xi_n$.
Since the usual condition for a Wronskian $w_0$ of two solutions
belonging to the same energy is $w_0=1$,
the natural
choice for $A$ is $A=-1$. In this case one recovers a discrete
analogue of the continuous Darboux transform
$(L^+\widehat \eta \,)_n=-\xi_n$.

So, we see that the discrete Darboux operators $L$ and $L^+$ realize
a one-to-one mapping of the spaces of solutions of discrete Schr\"odinger
equations (\ref{init}) and (\ref{fin}) for any $E\ne \lambda$.
Moreover, this mapping may be continued to $E=\lambda$ by putting
$\xi_n\longleftrightarrow \widehat \eta _n$ and
$\widehat \xi_n\longleftrightarrow \eta_n$ and considering this as a linear
mapping. We conclude hence that the knowledge of all solutions
of the initial equation (\ref{init}) provides the knowledge
of all solutions of the transformed equation (\ref{fin}) and,
in particular, one can get all "physical" solutions in the usual
quantum mechanical sense.
For this purpose one notices first
that the norm of the
vector
$\widetilde\psi=\left\{\right.\!\widetilde\psi_n\!\left.\right\}=L\psi =\left\{L\psi_n\right\}$
is proportional to the norm of the vector
$\psi=\left\{\psi_n\right\}$,
$\langle \widetilde\psi|\widetilde\psi \rangle =\sum_n \widetilde\psi^*_n\widetilde\psi_n
=\sum_n\psi^{*}_{n}(L^+L\psi)_n=
(E-\lambda )\sum_n\psi^{*}_{n}\psi_n=
(E-\lambda )\langle \psi|\psi \rangle $
where the use of the factorization property has been made. This
means that all "physical" solutions of (\ref{init})
with $E\ne \lambda $
are
mapped by $L$ onto "physical" solutions of (\ref{fin}). It is
evident that the inverse statement is also true but instead of
$L$ one has to use $L^+$. To find all "physical" solutions of
(\ref{fin}), it remains hence to analyze the sequences with
eigenvalue $E=\lambda $ forming a 2-dimensional space spanned by
$\left\{\eta_n\right\}$ and
$\left\{\right.\!\widehat \eta_n\!\left.\right\}$
which is not a difficult problem.

As a final remark of this section we note that the formulae
(\ref{Lpsin}) and (\ref{Ldag}) define operators $L$ and $L^+$ only
on finite elements from $H$ (finite linear combinations of basis
vectors $|n\rangle $). But using the technique of polar
decompositions of closed operators in terms of quasi-projectors
developed in \cite{SamJPhys} for a continuous basis, one can
extend this action such that the operators $L$ and $L^+$ become
mutually conjugated with respect to the inner product in $H$
with well-defined domains of definitions.

\section{Discrete supersymmetries of the Schr\"odinger
\protect{\newline}
equation}

The usual supersymmetry of the continuous Schr\"odinger equation
is based on factorization properties of Darboux transformation
operators (see e.g. \cite{BSTMF}). Let us find their counterparts
for the discrete equations (\ref{init}) and (\ref{fin}).
For this purpose, let us
consider the action of the superposition of $L$ and $L^+$ on a
ket-vector $|\psi \rangle $. Using (\ref{Lpsin}) and
(\ref{Ldag}) one finds
$$
L^+L|\psi \rangle =\sum_n(L^+L\psi )_n|n\rangle \,,
$$
\begin{equation}\label{LdLp}
(L^+L\psi )_n=A^{*}_nB_{n-1}\psi_{n-1}+
B^{*}_nA_{n+1}\psi_{n+1}+
(|A_n|^2+|B_n|^2)\psi_n
\end{equation}
where $\psi_n$ are Fourier coefficients of a vector $|\psi \rangle\in H$ over the
basis $|n\rangle $.
If for the sake of definiteness we assume that $\lambda <q_n$ for
all $n$, then from Eq. (\ref{u}) it follows that both
$a_n\xi_{n-1}/\xi_n<0$ and $a_{n+1}\xi_{n+1}/\xi_n<0$ which permits us
evaluate absolute values in (\ref{LdLp}). Finally, using once again the
equation for $u_n$ (\ref{u}), one gets
\begin{equation}\label{fac1}
(L^+L\psi )_n=|A|\,(\,[h_0-\lambda ]\psi )_n~.
\end{equation}
Similar calculations lead to another factorization
\begin{equation}\label{fac2}
(LL^+\widetilde\psi \,)_n=|A|\,(\,[h_1-\lambda]\widetilde\psi)_n~.
\end{equation}
It is necessary to note that these factorizations take place for
any sequences $\left\{\psi_n\right\}$ and
$\left\{\right.$$\widetilde\psi_n$$\left.\right\}$
and the superposition of operators $L$
and $L^+$ acting in the space of sequences $\ell ^{\,2}$ is defined in the regular
way $L^+L\psi_n=L^+(L\psi )_n=L^+\widetilde\psi_n=(L^+\widetilde\psi)_n$.
But when $\left\{\psi_n\right\}$ is an eigenelement of $h_0$,
$h_0\psi_n\equiv (h_0\psi )_n=E\psi_n$,
one gets from (\ref{fac1}) $(L^+L\psi )_n=(E-\lambda)\psi_n$
where we put $A=-1$. Similarly, when
$h_1\widetilde\psi_n\equiv (h_1\widetilde\psi)_n=E\widetilde\psi_n$
one obtains using (\ref{fac2})
$(LL^+\widetilde\psi \,)_n=(E-\lambda)\widetilde\psi_n$

We can now imitate supercharge operators by introducing nilpotent
difference matrix operators
\begin{equation}\label{QQ}
{\bbb Q}=\left(
\begin{array}{cc}
0&0\\
L&0
\end{array}
\right)\,,\quad
{\bbb Q}^+=\left(
\begin{array}{cc}
0&L^+\\
0&0
\end{array}
\right)
\end{equation}
acting in a vector space of 2-component column-vectors
\begin{equation}
\left\{\Psi_n\right\}=
\left(
\begin{array}{c}
\left\{\psi_n\right\} \\ \left\{\widetilde\psi_n\right\}
\end{array}
\right)\,,
\end{equation}
where $\psi_n$ and $\widetilde\psi_n$ are elements of arbitrary sequences
from $\ell^{\,2}$.
The superhamiltonian
\begin{equation}
{\bbb H}=
\left(
\begin{array}{cc}
h_0&0\\
0&h_1
\end{array}
\right)
\end{equation}
together with supercharges (\ref{QQ}) close a simplest
superalgebra
\begin{equation}\label{superalg}
\left[\, \bbb Q,\bbb H\,\right]=
\left[\, \bbb Q^+,\bbb H\,\right]=0\,,\quad
\bbb Q^2=(\bbb Q^+)^2=0\,,\quad
\left\{ \bbb Q,\bbb Q^+ \right\}=\bbb H-\lambda \bbb I
\end{equation}
where $\bbb I$ is the $2\times 2$ unity matrix. Here commutation
relations follow from the intertwining relation for $L$, $L^+$,
$h_0$ and $h_1$ and the anticommutation relation is an
implication of the factorization properties for these operators.
The discrete eigenvalue problem for the superhamiltonian
$\bbb H$ is nothing but the collection of either equations
(\ref{init}) and (\ref{fin}) in a single matrix equation.
Since the operators
$\bbb Q$ and $\bbb Q^+$
 from (\ref{superalg}) act in the space
$\ell^{\,2}$, we associate them with discrete supersymmetries
of the Schr\"odinger equation.

\section{New non-local exactly solvable potentials}

Now we shall apply the technique of discrete SUSY (or
equivalently discrete Darboux transformations) to generating
non-local exactly solvable potentials.

Let us consider the free particle Hamiltonian
$h_0=p_x^2=-d^2/dx^2$. Since the momentum operator may be
expressed in terms of the harmonic oscillator creation
$a^+=id/dx+ix/2$ and annihilation $a=id/dx-ix/2$ operators
$p_x=-(a+a^+)/2$, the Hamiltonian $h_0$ is a quadratic form of
$a$ and $a^+$, $h_0=(a+a^+)^2/4$. Therefore, the action of $h_0$
on the oscillator basis $|n\rangle $ which in coordinate
representation looks like
\begin{equation}\label{psinH}
\psi_n(x)= \langle x|n\rangle =
(-i)^n(n!2^n\sqrt{2\pi })^{-1/2} e^{-x^2/4}H_n(x/\sqrt 2)
\end{equation}
takes the form of a three term relation
\begin{equation}
h_0|n\rangle =\tfrac 14\sqrt{n(n-1)}|n-2\rangle +
\tfrac 14\sqrt{(n+1)(n+2)}|n+2\rangle +
(\tfrac n2+\tfrac 14)|n\rangle ~,
\end{equation}
where the use of the well-known properties of the creation and
annihilation operators
$a|n\rangle =\sqrt n|n-1\rangle $,
$a^+|n\rangle =\sqrt{n+1}|n+1\rangle $ has been made.
Let $|\psi_E\rangle $ be a continuous spectrum eigenket of $h_0$,
$h_0|\psi_E\rangle =E|\psi_E\rangle $. Then using the
self-adjointness of $h_0$ we get for the inner product
$\psi_n=\psi_n(E)=\langle \psi_E|n\rangle $ the following
discrete eigenvalue problem
\begin{eqnarray}
&a_n\psi_{n-2}+a_{n+2}\psi_{n+2}+q_n\psi_n=E\psi_n\,,
\label{init1}\\
&a_n=\tfrac 14\sqrt{n(n-1)}\,,\quad
q_n=(\tfrac n2+\tfrac 14)\label{anpn1}~.
\end{eqnarray}
A "physical" solution to this problem, $\psi_n=\psi_n(E)$,
can be easily obtained since it coincides with the
Fourier image  of the function (\ref{psinH})
\begin{equation}
\psi_n=2(n!2^n\sqrt{2\pi })^{-1/2}e^{-E}H_n(\sqrt {2E}\,)~.
\end{equation}
Here $H_n(z)$ are Hermite polynomials.
It is not difficult to convince ourselves that Eq.
(\ref{init1}) represents another form of the well-known
recursion relations for the Hermite polynomials.

The Hamiltonian $h_0$ is the kinetic energy operator for
$h\equiv h_1=h_0+V$.
Let us consider $V$ such that
\begin{equation}\label{Vnonloc}
V|n\rangle =d_n|n-2\rangle +
d_{n+2}|n+2\rangle +r_n|n\rangle
\end{equation}
In coordinate representation this operator is not reduced to the
multiplication on a function. We conclude therefore that this is
a non-local operator.

Let $|\widetilde\psi_E\rangle$ be an eigenvector of $h$,
$h|\widetilde\psi_E\rangle=E|\widetilde\psi_E\rangle$.
Then the eigenvalue problem for the inner products
$\widetilde\psi_n=\widetilde\psi_n(E)=\langle \widetilde\psi_E|\psi_n\rangle$
is just reduced to the discrete eigenvalue equation (\ref{fin})
which in our case has the form:
\begin{equation}\label{dnfi}
\widetilde a_n\widetilde\psi_{n-2}+\widetilde a_{n+2}\widetilde\psi_{n+2}+
\widetilde q_n\widetilde\psi_n=E\widetilde\psi_n\hphantom{d_n=\widetilde d_n+}
\end{equation}
where
\begin{equation}
\widetilde a_n=d_n+\tfrac 14\sqrt{n(n-1)}\,,\quad\label{tdnrn}
\widetilde q_n=r_n+\tfrac n2+\tfrac 14~.
\end{equation}
Consider now a subclass of potentials
(\ref{Vnonloc}) for which Eq. (\ref{dnfi}) coincides with the
Darboux transform of the Eq. (\ref{init1}). In this case the
functions $\widetilde a_n$ and $\widetilde q_n$ are defined by Eqs. (\ref{bn}) and
(\ref{qn})
\begin{eqnarray}\label{dn}
\widetilde a_n=[a_na_{n+2}\xi_{n-2}\xi_{n+2}]^{-1/2}/\xi_n\,,
\hphantom{r_n=q_{n+2}I}\\
\widetilde q_n=q_{n+2}-a_{n+2}{\xi_{n+2}}/{\xi_n}+
a_{n+4}{\xi_{n+4}}/{\xi_{n+2}}\label{rn}
\end{eqnarray}
where $a_n$ and $q_n$ are given in (\ref{anpn1}) and $\xi_n$ is a
particular solution of (\ref{init1}) at $E=\lambda <0$, i.e.
\begin{equation}
\xi_n=(n!\,2^n)^{-1/2}H_n(\sqrt{2\lambda})~.
\end{equation}
Note that these functions are real for $n$ even and purely
imaginary for $n$ odd and therefore $\widetilde a_n$ and $\widetilde q_n$ are real in
either case. Solutions of Eq. (\ref{dnfi}) are found with the aid
of Eq. (\ref{vfin})
\begin{equation}
\widetilde\psi_n=\left[ \frac{a_{n+2}}{\xi_n\xi_{n+2}} \right]^{1/2}
(\xi_n\psi_{n+2}-\xi_{n+2}\psi_n)
\end{equation}
which gives us continuous spectrum eigenfunctions for the
Hamiltonian $h=h_0+V$ with the interaction $V$ given in the form of
an infinite tridiagonal matrix
\begin{equation}
\langle k|V|n\rangle =
d_n\delta_{k,n-2}+ d_{n+2}\delta_{k,n+2}+r_n\delta_{k,n}
\end{equation}
with $d_{n}$ and $r_n$ defined with the aid of (\ref{tdnrn}),
(\ref{dn}) and (\ref{rn}).
It is not difficult to see that  $\eta_n$ found by
(\ref{vn}) behaves as ${n}^{-1/2}$ and  $\widehat \eta_n$
given by (\ref{tvn}) behaves as $n^{3/2}$ at large $n$. We
conclude hence that no discrete levels
in the potential $V$
is created by our method and this is a purely scattering
potential.

BFS acknowledges a partial support from the Russian
Foundation for Basic Researches.
AAS is grateful to Professors
H.V. Von Geramb and V. Inozemtzev for useful discussions.

\end{document}